\documentclass[preprint,12pt]{elsarticle}

\usepackage{graphicx}
\graphicspath{{fig/}}
\usepackage{amssymb}
\usepackage{amsmath}
\usepackage{lineno}

\journal{Optical Fiber Technology}

\begin{document}

\begin{frontmatter}


\title{Random Mode Coupling Assists Kerr Beam Self-Cleaning in a Graded-Index Multimode Optical Fiber}



\author[label1]{Oleg S. Sidelnikov\corref{cor1}}
\ead{o.s.sidelnikov@gmail.com}
\author[label1,label2]{Evgeniy V. Podivilov}
\author[label1,label3]{Mikhail P. Fedoruk}
\author[label1,label4]{Stefan Wabnitz}
\address[label1]{Novosibirsk State University, Novosibirsk 630090, Russia}
\address[label2]{Institute of Automation and Electrometry SB RAS, Novosibirsk 630090, Russia}
\address[label3]{Institute of Computational Technologies SB RAS, Novosibirsk 630090, Russia}
\address[label4]{Dipartimento di Ingegneria dell'Informazione, Elettronica e Telecomunicazioni, Sapienza Universit\`a di Roma, 00184 Roma, Italy}
\cortext[cor1]{Corresponding author}

\begin{abstract}
In this paper, we numerically investigate the process of beam self-cleaning in a graded-index multimode optical fiber, by using the coupled-mode model.  We introduce various models of random linear coupling between spatial modes, including coupling between all modes, or only between degenerate ones, and investigate the effects of random mode coupling on the beam self-cleaning process. The results of numerical investigations are in complete agreement with our experimental data.
\end{abstract}


\end{frontmatter}


\section{Intoduction}

Spatiotemporal light beam dynamics in multimode fibers (MMFs) has recently attracted a renewed interest, both from a fundamental physics perspective, and for its potential in various fields of practical application~\cite{Richardson2013,Richardson2010,Wright2015R31,Krupa2016}. Different intriguing nonlinear phenomena, such as multimode optical solitons~\cite{Renninger2013,Wright:15}, parametric instabilities leading to ultra-wideband dispersive wave sideband series generation~\cite{Wright2015R31,Wright2015}, geometric parametric instability~\cite{Longhi2007,Krupa2016}, supercontinuum generation~\cite{Lopez-Galmiche2016,KrupaLuot:16}, and self-induced beam cleaning~\cite{Krupa2016,Krupa2017,Liu2016,Wright2016}, to name a few, have only recently been experimentally observed in MMFs.

It is well known that when propagating along a multimode fiber, light is affected by random linear coupling caused by stress, bending, or technological irregularities of the fiber. This leads to a rapid distortion of the spatial intensity profle of the input beam, and to the appearance of highly speckled structures at the fiber output. Recent experiments in this area~\cite{Krupa2016,Krupa2017,Liu2016,Wright2016,PhysRevLett.122.103902} have demonstrated that the Kerr effect in a graded-index (GRIN) MMFs can overcome such speckle distortions, leading to a so-called spatial beam self-cleaning effect. As a result, one obtains a robust bell-shaped nonlinear beam, which at the fiber output has a diameter close to that the fundamental mode.

Physical experiments in nonlinear optics require complex and costly equipment, therefore the possibility of performing an accurate mathematical modeling of different optical systems is of paramount importance. In addition, analytical evaluations and numerical results often help us to confirm the results of experiments, and permit to select directions for further research. In this paper, we numerically investigate the process of Kerr beam self-cleaning in GRIN MMFs. Unlike previous approaches~\cite{Krupa2016,Krupa2017}, where the nonlinear propagation of a spatial beam along an optical fiber was described by the generalized 3D nonlinear Schrodinger equation, in this work we use the coupled-mode model~\cite{PhysRevLett.122.103902}. In order to reproduce the generation of output intensity speckles in the linear regime, caused by various imperfections of the fiber, we added a random linear coupling to the nonlinear coupled mode model. We consider various types of linear mode coupling, and investigate their effects on the beam self-cleaning process as reported in the recent experiments by different groups.

\section{Mathematical model}

We consider a GRIN fiber with the following refractive index profile:
\begin{equation*}
n(r) = 
\begin{cases}
n_{0} \sqrt{1-\Delta_\beta r^2} & r < a\\
n_{cl} & r \geq a
\end{cases},
\end{equation*}
where $n_{0}$ and $n_{cl}$ are the refractive indices of the core and cladding, respectively, $a$ is the core radius, and $\Delta_\beta = \sqrt{2\Delta}/a$ is the mode spacing, where $\Delta = (n_0^2-n_{cl}^2)/2n_0^2$.

Let us introduce the following dimensionless and normalized variables:
\begin{gather*}
\rho = \frac{r}{r_0}, \;\;\; r_0 = 1 / \sqrt{k \Delta_\beta}, \\
\zeta = z \Delta_\beta, \;\;\; A = \Psi \frac{\sqrt{P}}{r_0}, \\
p = \int{|A|^2 d^2 \vec{r}} / P_{sf} = (P/P_{sf}) \int{|\Psi|^2 d \vec{\rho}^2} = P / P_{sf},
\end{gather*}
where $A$ is the field amplitude, $k = 2\pi n_{co}/\lambda$ is the wave vector, $\lambda$ is the wavelength, and $P_{sf} = n_0 / 2n_2k \approx 1$~MW is the power threshold for catastrophic self-focusing. In this notation, the equation for the normalized field envelope reads as:
\begin{equation}
2 i \frac{\partial \Psi}{\partial \zeta} = \frac{\partial^2 \Psi}{\partial \rho^2} - \rho^2 \Psi + D \frac{\partial^2 \Psi}{\partial \zeta^2} + p |\Psi|^2\Psi,
\label{e:envelopeeq}
\end{equation}
where $D = \Delta_\beta / k$ is the angular dispersion coefficient. The complex field envelope in a GRIN fiber can be decomposed into a sum of spatial modes multiplied by the evolving mode amplitudes $A_{p,m}$:
\begin{equation}
\Psi \left(\zeta,\vec{\rho}\right) = \sum_{m,p = 0}^{\infty} A_{p,m}(\zeta) U_{p,m}(\vec{\rho}) \exp^{i(n+1)\zeta},
\label{e:fielddecomp}
\end{equation}
where
\begin{equation}
U_{p,m}(\vec{\rho}) = N_{p,m} \rho^{|m|} L_p^{|m|}\left(\rho^2\right) \exp^{-\rho^2/2} \exp^{im\phi}.
\label{e:modedistrib}
\end{equation}

Here $L_p^{|m|}$ are Laguerre polynomials, $N_{p,m}$ is a normalization coefficient, and $n = 2p + |m|$. By substituting the decomposition~\eqref{e:fielddecomp} into the equation~\eqref{e:envelopeeq}, and neglecting rapidly oscillating resonant four-wave mixing (FWM) terms, one obtains the coupled-mode model equations:

\begin{multline}
2i \frac{dA_{p,m}}{d\zeta} = D (n + 1)^2 A_{p,m} + \sum_{m_1,p_1} C_{p,p_1}^{m,m_1} A_{p_1,m_1} + \\ 
+ p \sum_{m+m_1 = m_2 + m_3} \sum_{p+p_1 = p_2 + p_3} f_{p,p_1,p_2,p_3}^{m,m_1,m_2,m_3} A_{p_1,m_1} A_{p_2,m_2} A_{p_3,m_3},
\label{e:ampliteq}
\end{multline}
where 
\begin{equation}
f_{p,p_1,p_2,p_3}^{m,m_1,m_2,m_3} = \int \int U_{p_1,m_1}^* U_{p_2,m_2} U_{p_3,m_3} U_{p,m}^* dxdy.
\end{equation}

These equations describe the evolution not of the full field, but separately the evolution of the amplitude of each mode $A_{p,m}$. This approach may significantly reduce the computation time, due to the use of a large integration step, which is permitted by neglecting rapidly oscillating FWM terms. To account for the various imperfections of the MMF, caused by production, bending or tilting, we added a linear coupling term to the equations. Coefficients $C_{p,p_1}^{m,m_1}$ correspond to random linear coupling between spatial modes. In order to preserve the total power, it is necessary that the matrix $C = \{C_{p,p_1}^{m,m_1}\}$ be Hermitian. Each element of the matrix $C$ is normally distributed with zero mean, and varies randomly during propagation along the fiber. In this contribution, we describe various models of random linear coupling between spatial modes, including coupling between all modes, or only between degenerate ones, and investigate the consequences of random mode coupling on the beam self-cleaning process.

Let us rewrite the system of equations~\eqref{e:ampliteq} in the matrix form:
\begin{equation}
\frac{dA}{d\zeta} = \left(- i \frac{D}{2}L - \frac{1}{2}C - i \frac{p}{2} NL\right)A = MA,
\label{e:matrixeq}
\end{equation}
where
\begin{gather}
L_{ii} = (n_i+1)^2, \\
{NL}_{ij} =  \sum_l \sum_k f_{p_l,p_k,p_i,p_j}^{m_l,m_k,m_i,m_j} A_{p_l,m_l} A_{p_k,m_k}.
\label{e:NLmatrix}
\end{gather}

Here, for the matrix NL in~\eqref{e:NLmatrix}, the summation passes by the same modes as in the equation~\eqref{e:ampliteq}. We numerically solve the propagation equations~\eqref{e:matrixeq} by using the the explicit finite-difference scheme:
\begin{equation}
\frac{A^{q+1} - A^{q-1}}{2h} = M^q A^q,
\end{equation}
where h is the integration step. The nonlinear coupling coefficients $f_{p,p_1,p_2,p_3}^{m,m_1,m_2,m_3}$ were determined by direct numerical calculation of the corresponding overlap integrals. 

In our investigation, we consider a 62.5 $\mu$m GRIN fiber with a core refractive index of 1.47 and $\Delta = 8.8 \cdot 10^{-3}$. At the fiber input, we suppose to launch a Gaussian beam of a certain radius at $1/e^2$, and at the wavelength of 1064 nm. Since we use a coupled-mode model, first we need to decompose the input beam into modes, thereby setting the initial modal power distribution, and obtaining the initial data for the equation~\eqref{e:ampliteq}. The considered spatial modes are orthonormal, so to find the decomposition coefficients, we need multiply the beam by the corresponding mode field distribution~\eqref{e:modedistrib} and integrate over $xy$ plane. In our simulations, we only consider spatial modes with mode number $n \leq 16$ (153 modes in total).

\begin{figure}[ht!]
	\begin{minipage}{0.49\linewidth}
		\raggedleft{\includegraphics[width=0.9\linewidth]{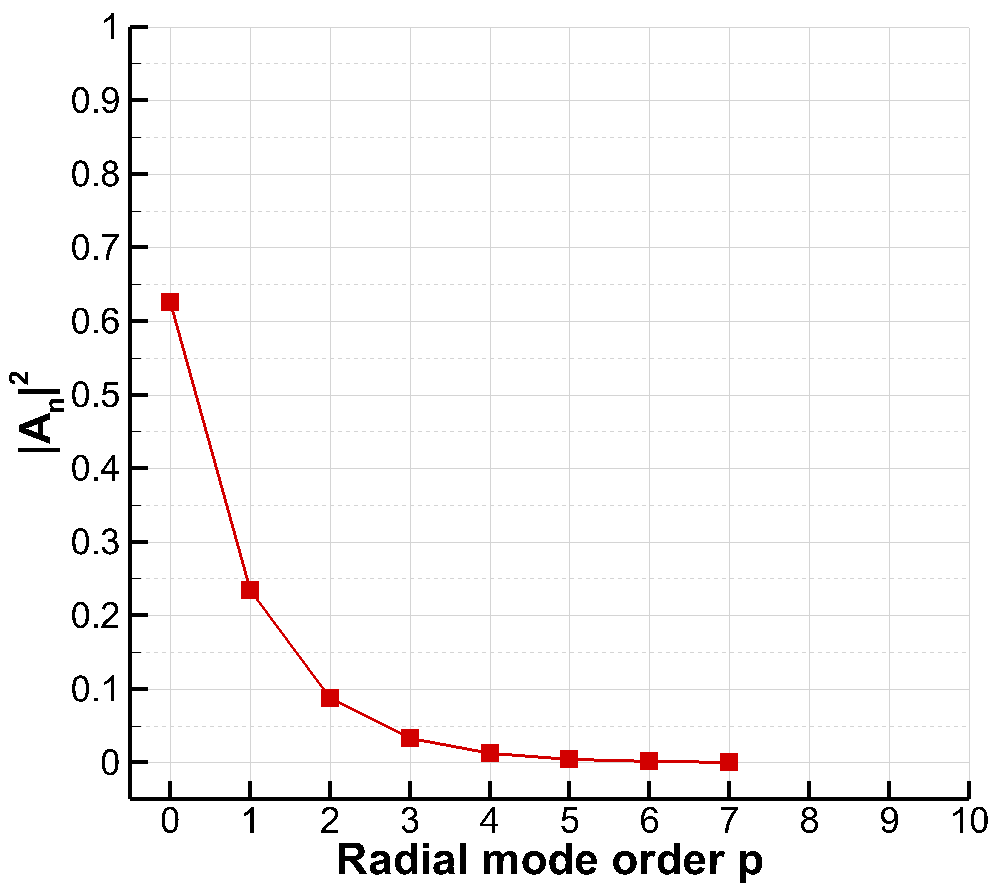}} a)
	\end{minipage}
	\hfill
	\begin{minipage}{0.49\linewidth}
		\raggedright{\includegraphics[width=0.9\linewidth]{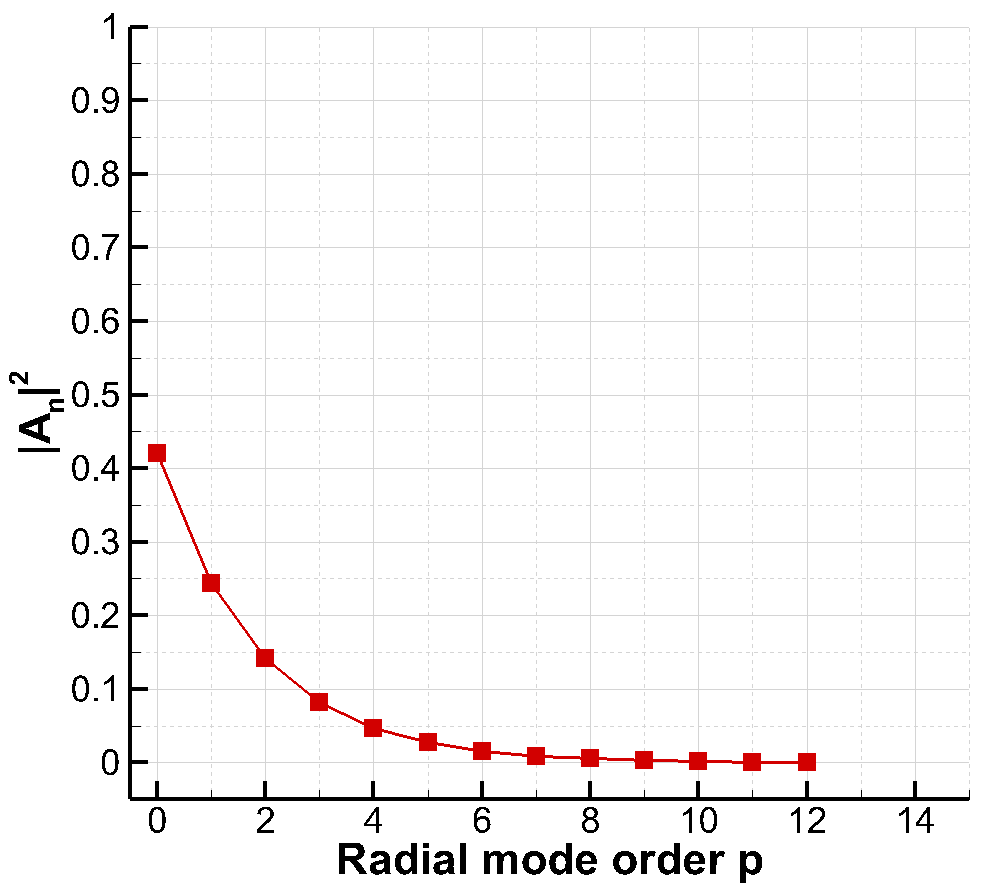}} b)
	\end{minipage}
	\caption{ Initial power distribution over the radial modes of Gaussian beams with radii of 15 (a) and 20 (b) microns.}
	\label{f:powerdistrib}
\end{figure}

Figure~\ref{f:powerdistrib} shows the initial power distribution over the radial modes of Gaussian beams with radii of 15 and 20 microns, obtained by using the procedure described above.

As we can see in Figure~\ref{f:powerdistrib}, with increasing radius of the input beam, the fraction of energy in the fundamental mode decreases, and more higher-order modes are excited.

\section{Numerical results}

We start out investigation by considering the model without any random linear coupling between modes, i.e., we set $(C_{p,p_1}^{m,m_1} = 0)$. Our purpose here is to show that we can reproduce the process of beam self-cleaning in the nonlinear regime using the considered coupled-mode model. In this case, the fundamental mode intensity increases up to a certain level during the propagation, and as a result we obtain a robust nonlinear beam, that at the fiber output has a size that is close to the fundamental mode.

\begin{figure}[h]
		\begin{minipage}{0.49\linewidth}
		\raggedleft{\includegraphics[width=0.9\linewidth]{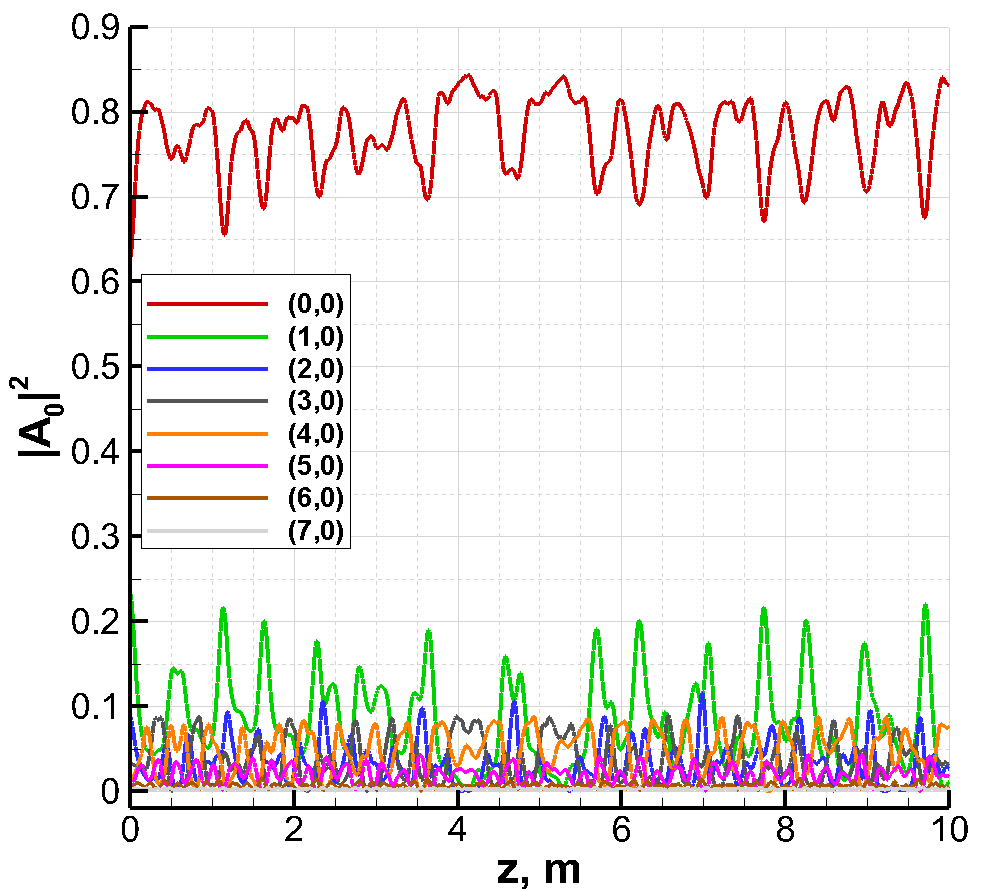}} a)
	\end{minipage}
	\hfill
	\begin{minipage}{0.49\linewidth}
		\raggedright{\includegraphics[width=0.9\linewidth]{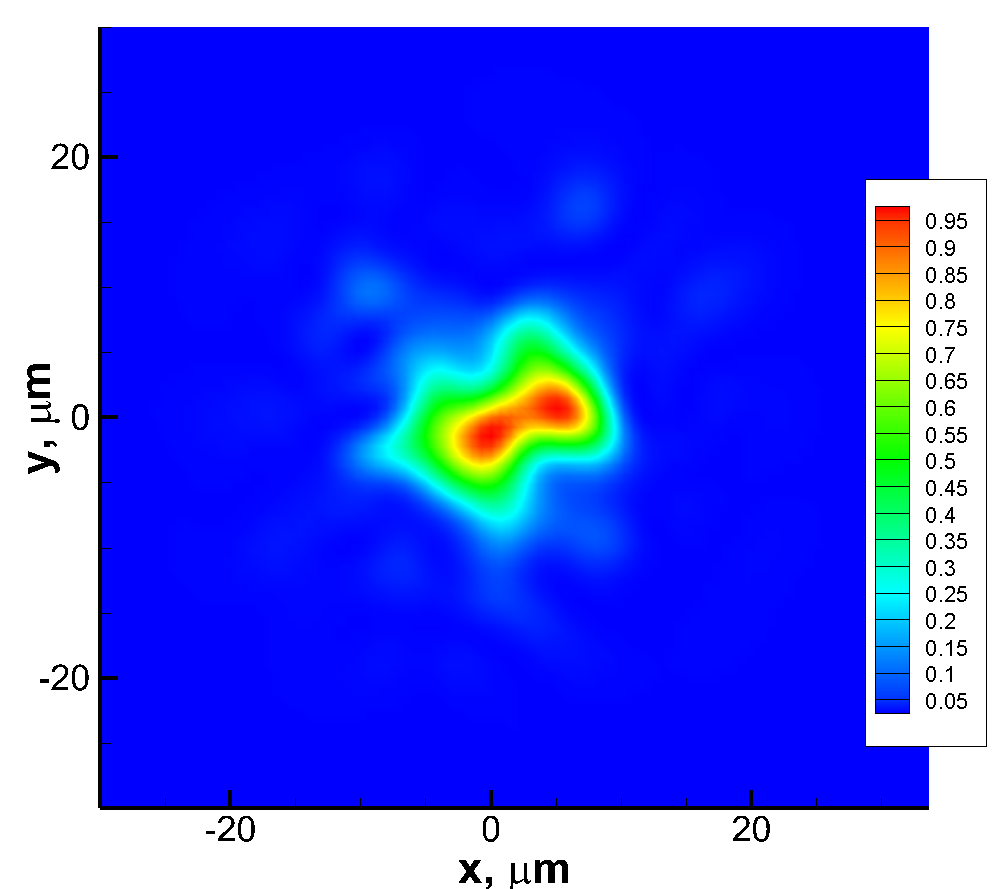}} b)
	\end{minipage}
	\caption{Dynamics of energy distribution by modes (a) and the output field (b) for the model without random linear coupling.}
	\label{f:endyn_wocoupl}
\end{figure}

Figure~\ref{f:endyn_wocoupl} shows the spatial dynamics of the energy distribution among the modes, as well as the output field for the initial beam with a radius of 15 $\mu$m and a power of 10 kW, corresponding to a model without no random linear coupling. In this case, in spite of being in a strong nonlinear regime, the power of the fundamental mode oscillates along the fiber, and we still obtain a speckled beam at the fiber output. In addition, in order to comply with the experiments, in the linear regime we must observe an highly speckled structure at the fiber output. But since we consider the model without linear coupling, the beam will not change during propagation in the linear regime, and the output field will be the same as the input field.

To obtain a speckled intensity pattern in the linear regime, in further research we used the full system of equations~\eqref{e:ampliteq} with a non-zero matrix $C$. First, we considered the case in which all spatial modes are linearly coupled to each other, and the coefficients $C_{p,p_1}^{m,m_1}$ are random variables that change with each integration step. This may correspond to the case of various fiber imperfections, resulting in random fluctuations in the refractive index of the core, or in the core diameter.

\begin{figure}[ht!]
	\centering\includegraphics[width=0.55\linewidth]{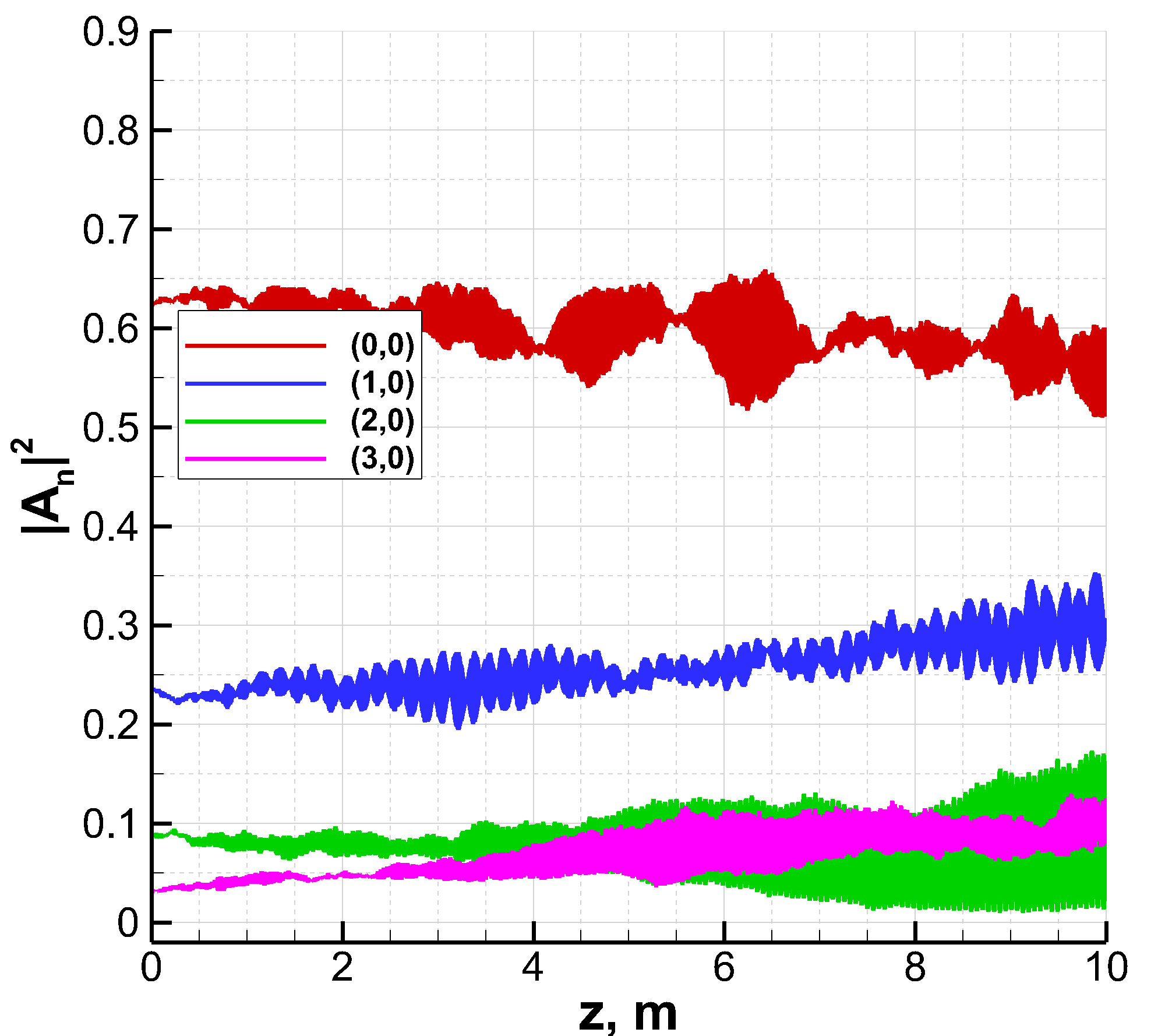}
	\caption{Dynamics of energy distribution by modes for the model with random linear coupling between all modes.}
	\label{f:endyn_noise}
\end{figure}

Figure~\ref{f:endyn_noise} shows the dynamics of energy distribution among the modes, for an initial beam with a radius of 15 $\mu$m and a power of 10 W for this model. We can see that with this type of coupling, the energy of each mode oscillates rapidly during the propagation, but on average it varies slightly. In addition, this model introduces some difficulties for solving the propagation equations numerically. Therefore, when using the model with random linear coupling between all spatial modes, we cannot observe a speckle pattern at the output of the fiber.

Next, we considered a model in which only modes, for which the conditions $n_1 = n \pm 1$ and $m_1 = m \pm 1$ are satisfied, are linearly coupled, and in which each element of the matrix C varies randomly during propagation along the fiber, with a correlation length of 10 cm. Such a model may correspond to the case of fiber tilting, in which energy can be transfered only among neighbouring modes in the linear regime.

\begin{figure}[ht!]
	\centering\includegraphics[width=0.5\linewidth]{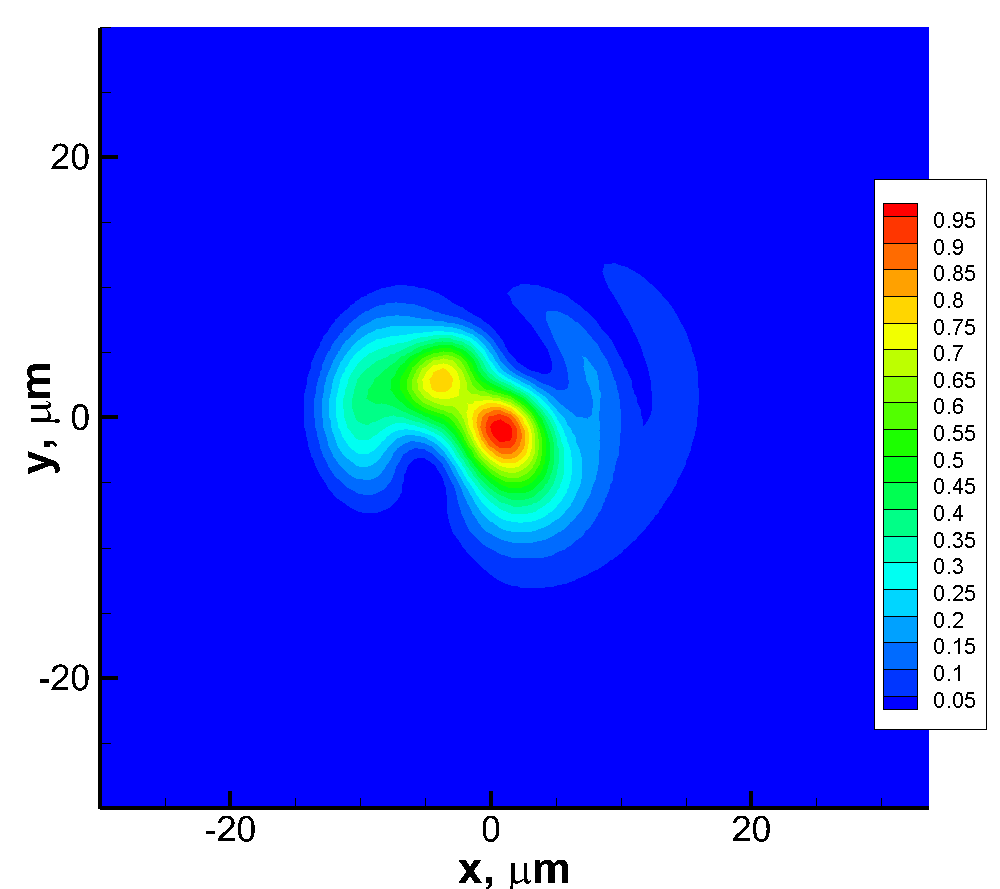}
	\caption{Output field for the model with random linear coupling between modes with $n_1 = n \pm 1$ and $m_1 = m \pm 1$.}
	\label{f:outfield_tilting}
\end{figure}

Figure~\ref{f:outfield_tilting} shows the output field for an initial beam with a radius of 15 $\mu$m, and a power of 10 W for the considered model. The beam at the fiber output is distorted, but it still does not appear to reproduce the highly speckled field which is obtained in the experiment. In addition, when using this type of coupling in the nonlinear regime, rapid oscillations of the fundamental mode power are observed, as in the case of the model without any random linear coupling.

Finally, we considered a model with random linear coupling between spatial modes with equal mode numbers $n$ only. In this case, only degenerate modes are linearly coupled, and the energy from the fundamental mode flows into azimuthal modes with the same number.

\begin{figure}[hb!]
	\begin{minipage}{0.49\linewidth}
		\raggedleft{\includegraphics[width=0.9\linewidth]{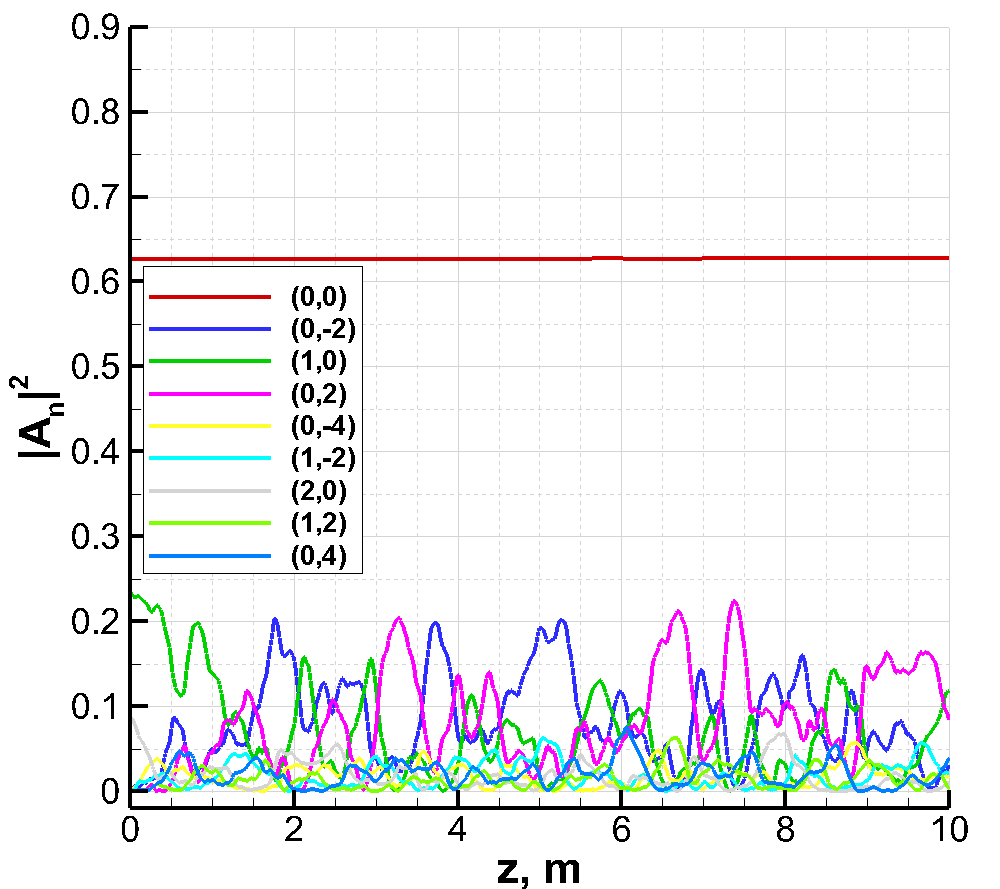}} a)
	\end{minipage}
	\hfill
	\begin{minipage}{0.49\linewidth}
		\raggedright{\includegraphics[width=0.9\linewidth]{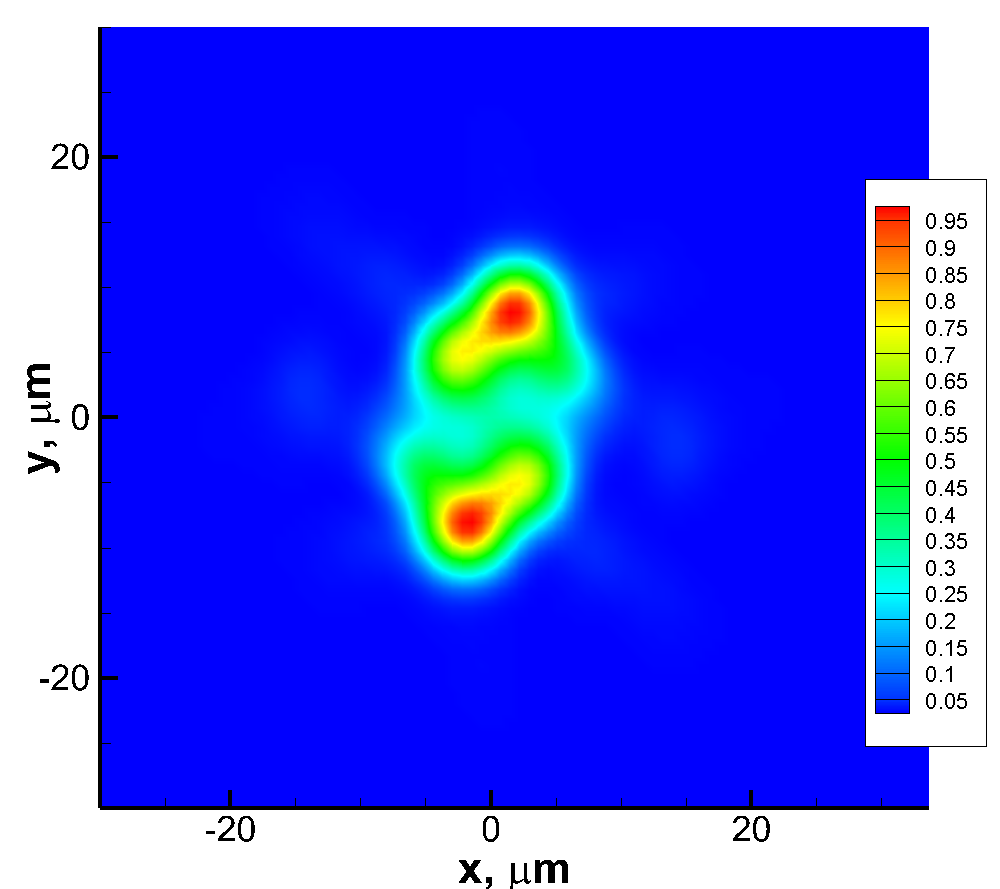}} b)
	\end{minipage}
	\caption{Dynamics of energy distribution by modes (a) and the output field (b) for the model with random linear coupling between modes with equal mode numbers in the linear regime.}
	\label{f:endyn_azimuth}
\end{figure}

\begin{figure}[ht!]
	\centering\includegraphics[width=0.5\linewidth]{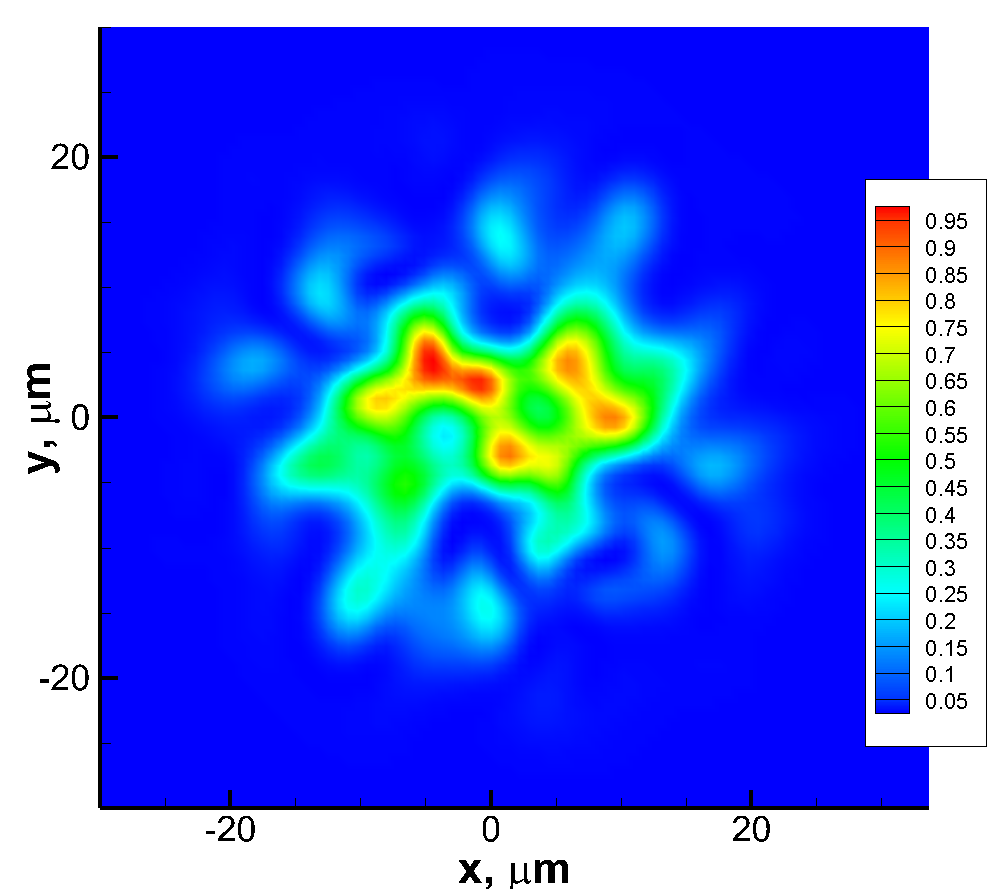}
	\caption{Output field for the model with random linear coupling between modes with equal mode numbers in the linear regime and for a spatially shifted initial beam.}
	\label{f:outfield_azimuthshift}
\end{figure}

Figure~\ref{f:endyn_azimuth} shows the spatial dynamics of the energy distribution among the modes and the output field for an initial beam with a radius of 15 $\mu$m, and a power of 10 W. In this case, we can see an intense exchange of energy between the modes, while the power of the fundamental mode does not change, since this is the only mode with $n = 0$.  As we can see, the output beam exhibits a speckle pattern, which however turns out to be symmetrical with respect to the center of the fiber. This can be explained by the fact that, if we launch an initial beam exactly into the centre of the fiber, then it decomposes only into radially symmetric modes. As a consequence, during subsequent propagation only azimuthal modes with even mode numbers are excited, because there are no radial modes among modes with odd numbers. But if we only slightly shift the beam centre, then also some of the azimuthal modes with odd mode numbers will be initially excited. In this case, we do obtain a more realistic-looking random speckled output field (see figure~\ref{f:outfield_azimuthshift}, for, e.g., an initial beam that is laterally shifted by 5 microns).

\begin{figure}[ht!]
	\begin{minipage}{0.49\linewidth}
		\raggedleft{\includegraphics[width=0.9\linewidth]{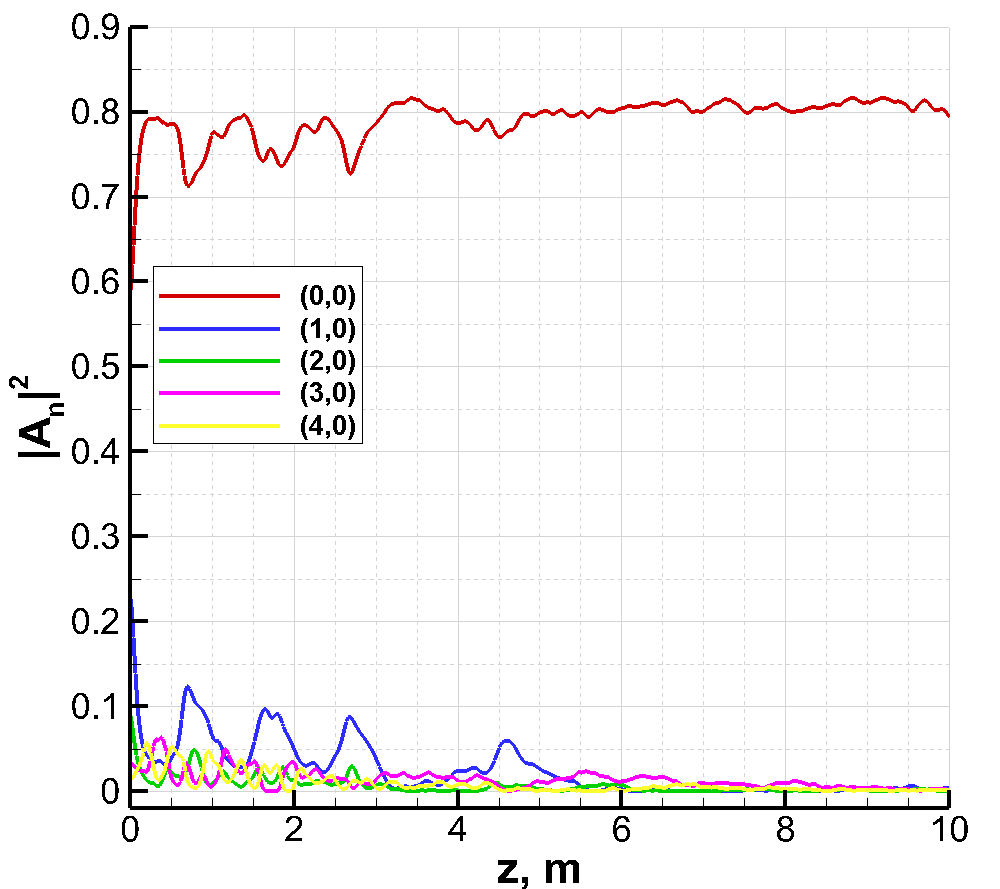}} a)
	\end{minipage}
	\hfill
	\begin{minipage}{0.49\linewidth}
		\raggedright{\includegraphics[width=0.9\linewidth]{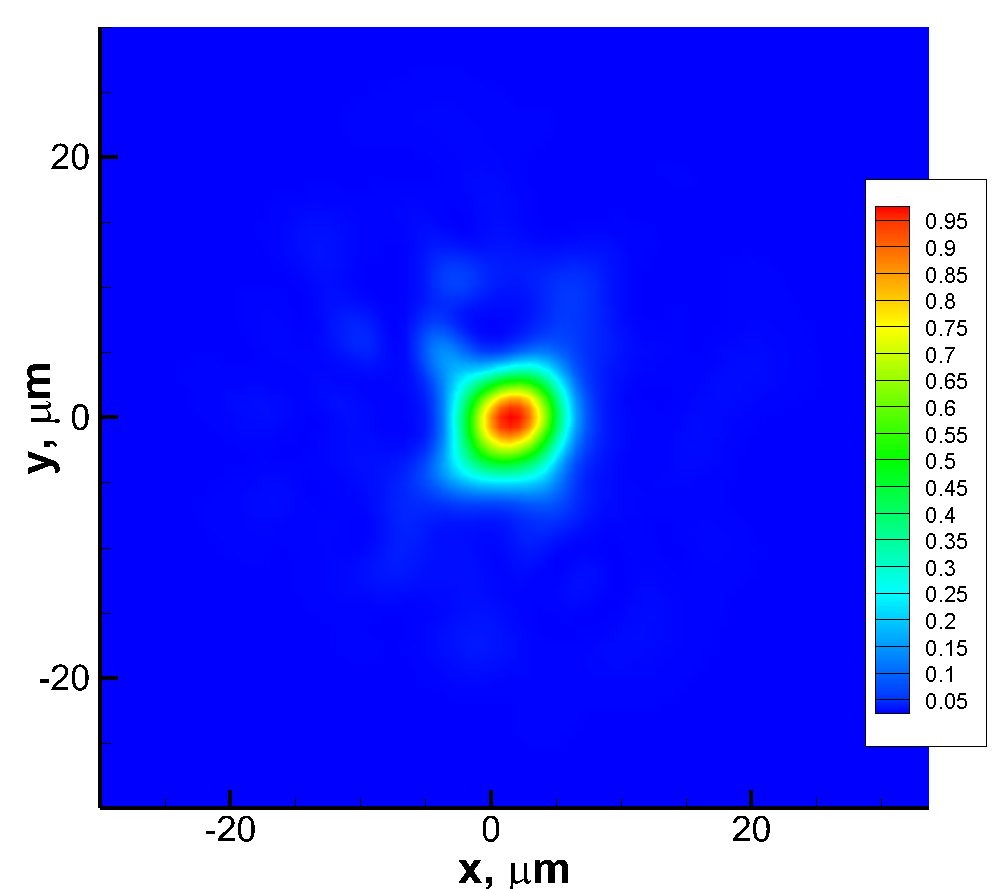}} b)
	\end{minipage}
	\caption{Dynamics of energy distribution by modes (a) and the output field (b) for the model with random linear coupling between modes with equal mode numbers in the nonlinear regime and for the shifted initial beam.}
	\label{f:endyn_azimuthshift_nonling}
\end{figure}

Figure~\ref{f:endyn_azimuthshift_nonling} shows the spatial dynamics of the energy distribution among modes, and the output field, for an initial beam with a radius of 15 $\mu$m, and a power of 10 kW, laterally shifted by 5 microns, corresponding to a numerical solution of the coupled mode model with random linear coupling between modes with equal mode numbers only.

In this case, one readily observes the appearance of a self-cleaning effect, and the fundamental mode power swiftly stabilizes upon propagation. The output beam for the considered model has a size that is close to that of the fundamental mode. It should be noted that the results of our numerical investigations are in complete agreement with our experimental data~\cite{PhysRevLett.122.103902}.

\section{Conclusion}

In this paper, we numerically investigated the process of beam self-cleaning in graded-index multimode optical fibers, by using the coupled-mode model. To account for the various imperfections of the fiber, we added a random linear coupling term to the model, and considered various options for its implementation. We investigated the effect of random coupling on beam self-cleaning in the nonlinear regime, and the occurrence of a speckled output field in the linear one. It was found that the results obtained using the model with a random linear coupling between modes with equal mode numbers are in complete agreement with available experimental data.

\section*{Acknowledgements}
This work was supported by the Russian Ministry of Science and Education (Grant 14.Y26.31.0017), and the European Research Council (ERC) under the European Union’s Horizon 2020 research and innovation programme (Grant No. 740355). The work of Oleg S. Sidelnikov was carried out within the framework of the State Research Task of the Ministry of Education and Science of the Russian Federation (Project No. 1.6366.2017/8.9).
\bibliographystyle{model1-num-names}
\bibliography{BSCN}

\end{document}